\documentclass[12pt]{article}

\usepackage{amssymb,amsmath}
\usepackage{subfig}
\usepackage{graphicx}
\usepackage{color}
\usepackage{hyperref}
\usepackage{listings}
\usepackage{siunitx} 
\usepackage{soul}

\usepackage{tikz}
\usepackage[export]{adjustbox}
\usepackage{amsmath}
\usepackage[normalem]{ulem} 

\usepackage[left=2cm,right=2cm,top=2cm,bottom=2cm]{geometry}
\usepackage{authblk}
\usepackage[doublespacing]{setspace}
\usepackage{multicol}
\usepackage{flafter}
\usepackage{bbold}
\usepackage{mathrsfs}
\DeclareMathOperator{\sinc}{sinc}
\usepackage{threeparttable} 
\usepackage{mathtools}

\usepackage{booktabs} 
\AtBeginDocument{
\heavyrulewidth=.08em
\lightrulewidth=.05em
\cmidrulewidth=.03em
\belowrulesep=.65ex
\belowbottomsep=0pt
\aboverulesep=.4ex
\abovetopsep=0pt
\cmidrulesep=\doublerulesep
\cmidrulekern=.5em
\defaultaddspace=.5em}

\usepackage[style=nature]{biblatex}
\hypersetup{
    unicode=false,          
    pdftoolbar=true,        
    pdfmenubar=true,        
    pdffitwindow=false,     
    pdfstartview={FitH},    
    pdfauthor={Kevin Multani},     
    colorlinks=true,       
    linkcolor=blue,          
    citecolor=red,        
}
\graphicspath{{figures/}}

\newcommand{\ket}[1]{$\left|#1\right>$}

\definecolor{dkgreen}{rgb}{0,0.6,0}
\definecolor{gray}{rgb}{0.5,0.5,0.5}
\definecolor{mauve}{rgb}{0.58,0,0.82}

\begin{document}
\title{Nondestructive detection of photonic qubits}
\author[1]{D. Niemietz$^{*}$}
\author[1,2]{P. Farrera}
\author[1]{S. Langenfeld}
\author[1]{G. Rempe}
\affil[1]{Max-Planck-Institute für Quantenoptik, Hans-Kopfermann-Str. 1, 85748 Garching, Germany}
\affil[2]{ICFO-Institut de Ciencies Fotoniques, The Barcelona Institute of Science and Technology, 08860 Castelldefels (Barcelona), Spain}

\date{\today}
\maketitle
\textbf{One of the biggest challenges in experimental quantum information is to keep the fragile superposition state of a qubit alive \cite{Bennett2000}. Long lifetimes can be achieved for material qubit carriers as memories \cite{Zhong2015}, at least in principle, but not for propagating photons that are rapidly lost by absorption, diffraction or scattering \cite{Gisin2007}. The loss problem can be mitigated with a nondestructive photonic qubit detector that heralds the photon without destroying the encoded qubit. Such detector is envisioned to facilitate protocols where distributed tasks depend on the successful dissemination of photonic qubits \cite{Sangouard2012, Boon2015}, to improve loss-sensitive qubit measurements \cite{Sciarrino2012, Gisin2010}, and to enable certain quantum key distribution attacks \cite{Sanders2000}. Here we demonstrate such a detector based on a single atom in two crossed fibre-based optical resonators, one for qubit-insensitive atom-photon coupling, the other for atomic-state detection \cite{Rempe2015}. We achieve a nondestructive detection efficiency upon qubit survival of $(79\pm3)\,\%$, a photon survival probability of $(31\pm1)\,\%$, and preserve the qubit information with a fidelity of $(96.2\pm0.3)\,\%$. To illustrate the potential of our detector we show that it can provide, already with current parameters, an advantage for long-distance entanglement and quantum-state distribution, resource optimization via qubit amplification, and detection-loophole-free Bell tests.}\\

\noindent The qubit is the elementary information unit in quantum information science \cite{Bennett2000}. Encoding the qubit into two modes of a single photon allows the distribution of quantum information over long distances. This has enabled a spectrum of experiments, ranging from fundamental tests of quantum physics \cite{Hensen2015, Giustina2015} to applications related to quantum communication \cite{Gisin2007} and quantum networks \cite{Rempe2015, Pan2018}. However, unavoidable absorption, diffraction and scattering losses of long transmission channels severely limit the transfer distance. It is important to emphasize that these losses typically occur independent of the state of the qubit that is encoded in the two optical modes, be it time bins or light polarization. In fact, in optical fibres the loss rate of the qubit carrier, the photon, can be many orders of magnitude larger than all decoherence rates of the encoded qubit. The situation is complementary for material qubit carriers that are hardly lost in any quantum information protocol demonstrated so far. For photons, the loss is fundamental and cannot be eliminated in any of the envisioned quantum information processing tasks. However, the loss effect could be mitigated by tracking the photon without destroying the encoded qubit, provided a nondestructive photonic qubit detector (NPQD) is available.

\begin{figure*}
\centering
\includegraphics[width=\linewidth]{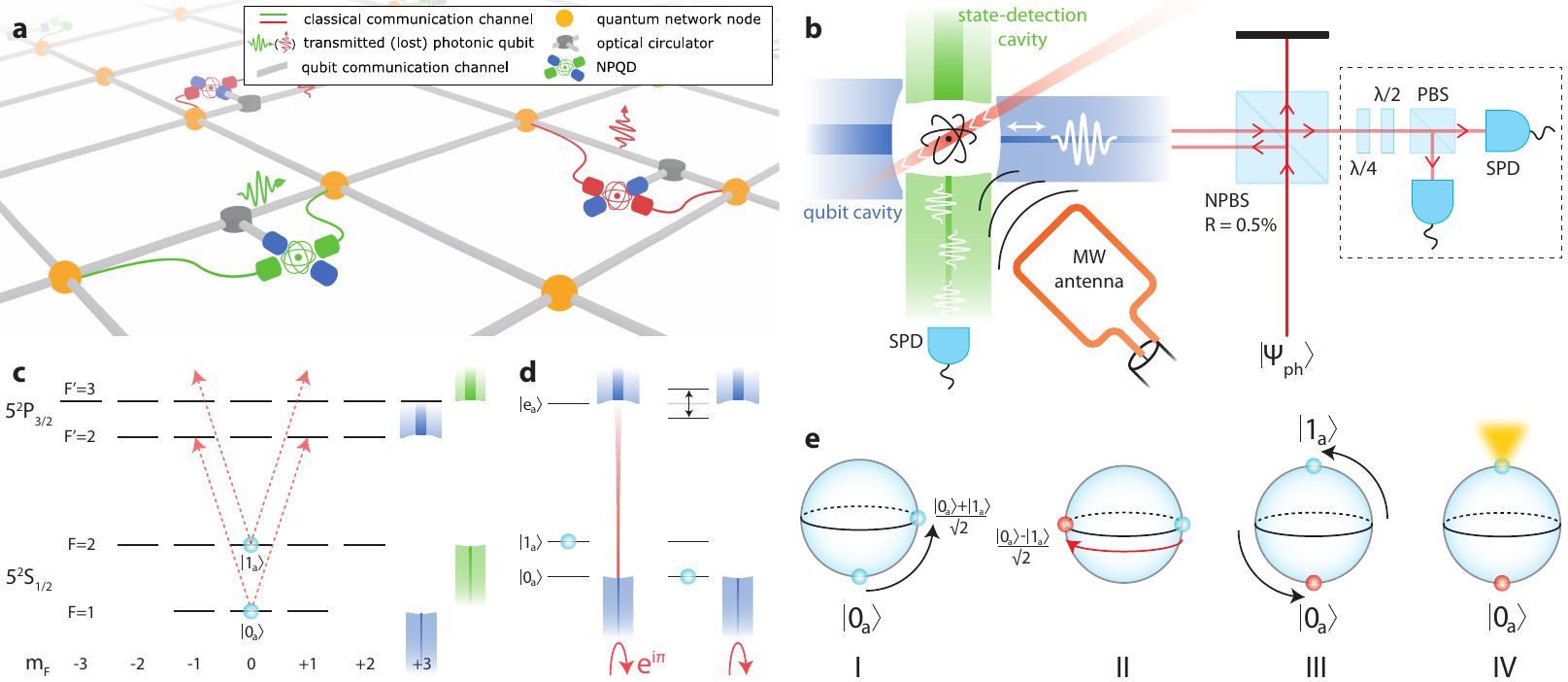}
\caption{\textbf{Nondestructive photonic qubit detector.} \textbf{a}, A quantum network including NPQDs. (Un)detected qubit photon events are colored in (red) green. \textbf{b}, The NPQD setup showing the crossed optical fibre cavities together with the microwave (MW) antenna and the state-detection beam (red solid line). Input photonic qubits (red arrows) are reflected by the NPQD via a highly transmittive non-polarizing beam splitter (NPBS) and later sent to a qubit measurement setup (black dashed rectangle) including $\lambda/2$ and $\lambda/4$ waveplates, a polarizing beam splitter (PBS) and single photon detectors (SPD). \textbf{c}, Atomic level scheme together with the qubit coupling transitions (dashed arrows). \textbf{d}, The atomic state affects the entrance of the qubit photon into the qubit cavity. \textbf{e}, Bloch spheres I--IV show the atomic state for different stages of the ideal NPQD scheme.} 
\label{fig:comic}
\end{figure*}

Once at hand, such a NPQD could communicate to both the sending and the receiving nodes of a quantum network whether a photonic qubit has been lost along the way or not (Fig.\,\ref{fig:comic}a). This loss monitoring has several advantages: First, it allows quantum communication schemes to stop the execution of further operations and restart the protocol when the qubit carrying photon was found to be lost along the communication channel. This is important when these operations are time-expensive or involve the use of precious resources such as long-distance entanglement for quantum teleportation. Secondly, the detector can precertify the arrival of a photonic qubit immediately before a planned quantum measurement so that the latter is only conducted when the qubit is present at the measurement input. This allows performing loss-sensitive qubit measurements even in the presence of high transmission losses. Examples are measurements limited by the detection noise, photon-receiving repeater schemes \cite{Loock2019}, or experiments where qubit loss leads to loopholes as in Bell tests \cite{Gisin2010, Sciarrino2012}. The latter are of key importance for device-independent quantum key distribution \cite{Scarani2009}.

Despite recent progress on the quantum non-demolition detection of optical and microwave fields \cite{Grangier1998, Guerlin2007, Johnson2010, Reiserer2013, Kono2018, Besse2018, Lee2019}, the nondestructive detection of the photonic energy without projection of the encoded qubit information remains an outstanding challenge. Some work related to the experimental implementation of NPQDs in the optical domain has been conducted, such as the nondestructive detection of bright two-mode light pulses using cross-phase modulation \cite{Sinclair2016}. This scheme still operates far away from the single-photon qubit regime. Another approach makes use of parametric down-conversion of the incoming qubit photon \cite{Jennewein2016}, where one down-converted photon heralds the qubit presence and the other provides the qubit information. This approach shows a severely limited efficiency ($-76\,$dB) due to the small conversion efficiency of the applied photon splitting process. Other related work includes heralded qubit amplifiers which use two ancilla photons that interfere with the input signal \cite{Pryde2013}. However, interference requires prior knowledge about pulse shape and arrival time and therefore restricts the range of possible applications. Finally, our latest work about a heralded quantum memory \cite{Brekenfeld2020} can, in combination with readout, be used to herald photonic qubits. But the single-photon herald signal is highly sensitive to losses and therefore causes a small detection efficiency. It is important to point out that (in contrast to the NPQD reported here) all mentioned qubit heralding schemes destroy the original incoming photon waveform that is important in applications where synchronisation and interference matters, as for time-bin qubits or in linear optical quantum computing \cite{Knill2001}.

We here demonstrate the implementation of a NPQD using a single atom strongly coupled to the modes of two independent optical resonators. Photonic polarization qubits are sent onto the qubit cavity where they are reflected while imprinting a $\pi$ phase shift on an atomic superposition state. By coherent manipulation of the atomic superposition state, phase information is mapped into population information which can be read out using the state-detection cavity. This signal heralds the nondestructive detection of the photonic qubit reflected off the qubit cavity. We show that the strong atom-photon interaction provided by both cavities allows for an efficient NPQD.  We also proof that the reflection mechanism preserves the temporal waveform of the light pulse, which is of key importance for photonic time-bin qubits.

The qubit-atom interaction mechanism is essentially a single-rail version of the scheme proposed by Duan \textit{et al.} \cite{Duan2004}. In our case, it starts with a $^{87}$Rb atom prepared in state $(\left|0_\text{a}\right>+\left|1_\text{a}\right>)/\sqrt{2}$ (Bloch sphere I in Fig.\,\ref{fig:comic}e). Atomic state \ket{0_\text{a}} is strongly coupled to optical cavity modes, thus preventing a photon to enter the cavity. Instead, the photon reflects from the cavity input mirror (Fig.\,\ref{fig:comic}d) so that \ket{\Psi_\text{ph}}\ket{0_\text{a}}$\rightarrow$+\ket{\Psi_\text{ph}}\ket{0_\text{a}} where \ket{\Psi_\text{ph}} denotes a photonic polarization qubit. In contrast, state \ket{1_\text{a}} owns no transition in resonance with the qubit cavity. Hence, the qubit photon enters the cavity and acquires a $\pi$ phase shift upon reflection, \ket{\Psi_\text{ph}}\ket{1_\text{a}}$\rightarrow 
-$\ket{\Psi_\text{ph}}\ket{1_\text{a}}. Together, the qubit photon flips the atomic superposition state to $(\left|0_\text{a}\right>-\left|1_\text{a}\right>)/\sqrt{2}$ (Bloch sphere II). Finally, a $\pi/2$ pulse rotates the atomic state to \ket{0_\text{a}} (\ket{1_\text{a}}) in case of one (no) qubit photon (Bloch sphere III), allowing to witness the atomic phase flip by means of deterministic atomic-state detection (Bloch sphere IV). 

Compared to previous experiments \cite{Reiserer2013}, our new scheme is made possible by two key ingredients: First, we choose state $\left|0_\text{a}\right>\coloneqq\left|F=1, m_F=0\right>$ so that the qubit cavity with its two left and right circularly polarized and frequency-degenerate eigenmodes couples strongly to \textit{two} atomic transitions, $\left|0_\text{a}\right>\leftrightarrow$ \ket{F'=2, m_F=\pm1} (Fig.\,\ref{fig:comic}c). The two coupling strengths are equal by definition but much smaller than for a cycling transition. We compensate this reduction by using a miniaturized fibre resonator \cite{Hunger2010}, achieving a coupling constant $g=2\pi\times\left(18.6\pm0.5\right)$\,MHz and a cooperativity of $C=1.67\pm0.09$ (see Methods). Second, the newly added state-detection cavity is tuned to the atomic cycling transition \ket{F=2} $\leftrightarrow$ \ket{F'=3}, enabling a strong fluorescence signal to be observed from $\left|1_\text{a}\right>\coloneqq\left|F=2, m_F=0\right>$. This is used to distinguish the two atomic states \ket{0_\text{a}} and \ket{1_\text{a}} with a fidelity of $(98.2\pm0.2)\,\%$ at a photon-number threshold of one.

The physical elements needed to implement the NPQD are shown in Fig.\,\ref{fig:comic}b. The photon qubit \ket{\Psi_\text{ph}} first passes a highly transmitting non-polarizing beam splitter (NPBS) and is then coupled into a single-mode optical fibre that connects to the qubit fibre cavity where the nondestructive interaction takes place. After reflection, the qubit photon is characterized by state tomography. Coherent manipulation of the atomic ground states is achieved by means of microwave (MW) fields emitted by an antenna close to the fibre cavities. The MW flips the atomic state in about $12\,\mu$s with a probability of $(97\pm1)\,\%$ (see Methods). The detector herald signal is generated by a laser that, together with the state-detection cavity, induces cavity-enhanced fluorescence \cite{Rempe2015} for typically $7.5\,\mu$s. The fluorescence photons are predominantly emitted into the cavity from where they are guided towards a single photon detector (SPD). Note that both fibre cavities are single-sided (details in \cite{Brekenfeld2020}), so that the dominant escape channels for cavity photons are the two single-mode fibres.

\begin{figure}
\centering
\includegraphics[width=0.5\linewidth]{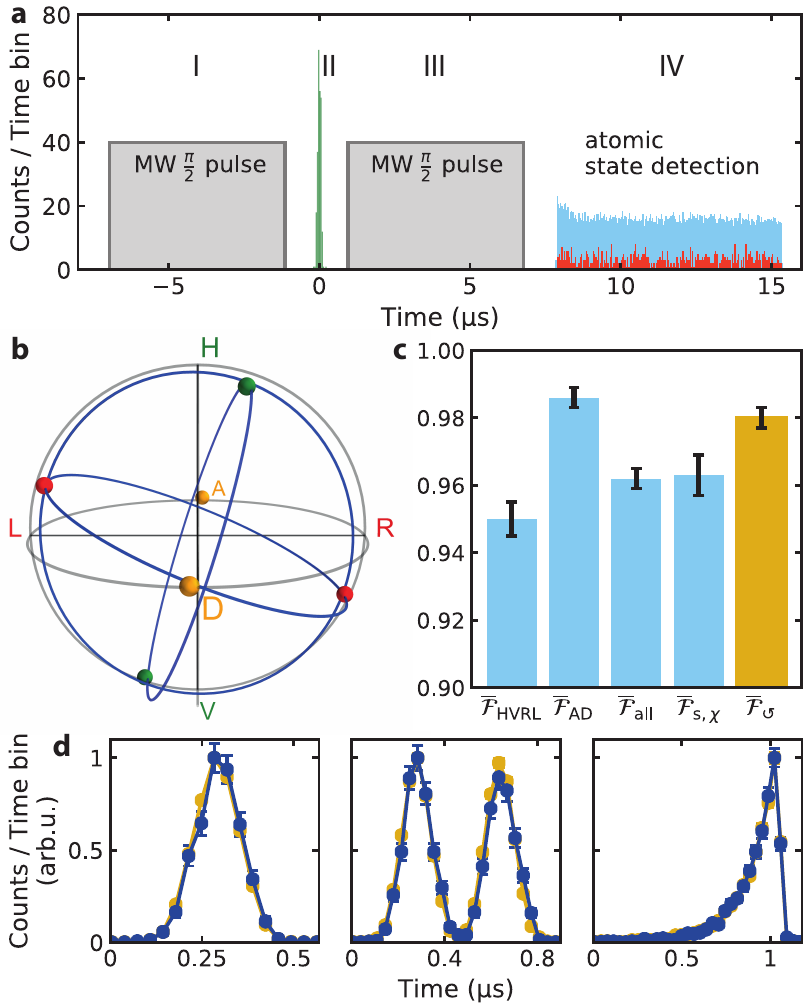}
\caption{\textbf{Experimental NPQD implementation.} \textbf{a}, Time histogram showing the NPQD sequence, which includes the microwave pulses, the reflected qubit photon events (green) and the state-detection photon events. Red (blue) colored state-detection data is (un)conditioned on a reflected photonic qubit detection. Labels I--IV refer to the Bloch spheres in Fig.\,\ref{fig:comic}e. \textbf{b}, Poincar\'{e} sphere showing the result of state tomography (coloured spheres) for reflected photonic qubits that were nondestructively detected. Labels indicate input polarizations. \textbf{c}, Bar chart of mean fidelities $\overline{\mathcal{F}}_\text{x}$ and the quantum process state fidelity $\overline{\mathcal{F}}_{\text{s},\chi}$. The Poincar\'{e} sphere shows a rotation around the $A/D$ axis which, if eliminated, leads to $\overline{\mathcal{F}}_\circlearrowleft$. \textbf{d}, The preservation of the photonic waveform is tested by reflecting photonic qubits with three different waveforms off the NPQD (blue points). Yellow points represent empty cavity measurements as reference. The error bars in \textbf{c} (\textbf{d}) describe the $1\sigma$ confidence interval (standard deviation). The error bar of $\overline{\mathcal{F}}_{\text{s},\chi}$ is assessed with a Monte Carlo method and represents the standard error.} 
\label{fig:FOM2}
\end{figure}

We now characterize the nondestructive detection of photonic polarization qubits. For this, weak coherent pulses at the single-photon level were used. Fig.\,\ref{fig:FOM2}a illustrates the NPQD temporal sequence by showing a time histogram of detected SPD counts at the qubit measurement setup and the state-detection cavity output. The mean photon number here is $\left|\alpha\right|^2=0.13$ in front of the qubit cavity. As can be seen, the reflection of photonic qubits at the NPQD (green bars) has a big effect on the number of observed state-detection photons (red bars) compared to the situation where no qubits are sent (blue bars). This enables an efficient detection of the photonic qubits. 
However, the definition of the NPQD efficiency is ambiguous as it depends on the use case. Relevant parameters that relate to the NPQD efficiency include the conditional probability of detecting the atom in the qubit-heralding state \ket{0_\text{a}}, either given a qubit at the qubit cavity output ($1_\text{oq}$), $P\left(0_\text{a}|1_\text{oq}\right)=(79\pm3)\,\%$, or given a qubit at the NPQD input ($1_\text{iq}$), $P\left(0_\text{a}|1_\text{iq}\right)=(45\pm2)\,\%$, and the photon survival probability $\eta_{\rm{surv}}=(31\pm1)\,\%$. Another figure of merit for any detector is the dark-count probability which is $p_\text{DC}=(3.3\pm0.2)\,\%$ for our NPQD.

As NPQDs should preserve the qubit, polarization state tomography was conducted on the outgoing photonic qubits and compared with the incoming qubits. With this we calculate the fidelity $\mathcal{F}=\left<\Psi_\text{in}\right|\rho_\text{out}\left|\Psi_\text{in}\right>$. Six orthogonal polarization states were tomographically investigated (Fig.\,\ref{fig:FOM2}b), showing an overall mean state fidelity of $\overline{\mathcal{F}}_\text{all}=(96.2\pm0.3)\,\%$ conditioned on a nondestructive qubit detection. A major contribution to the infidelity comes from a polarization rotation around the $A/D$ axis, attributed to a residual birefringence of the qubit cavity (see Methods). The mean fidelity for (non-) rotated states is given by $\left(\overline{\mathcal{F}}_\text{AD}\right)$ $\overline{\mathcal{F}}_\text{HVRL}$ (Fig.\,\ref{fig:FOM2}c). However, taking into account this rotation in the calculation (which would be experimentally feasible by placing retardation wave plates after the qubit cavity) leads to an overall mean state fidelity of $\overline{\mathcal{F}}_\circlearrowleft=(98.0\pm0.3)\,\%$ (orange bar). The remaining $2\,\%$ infidelity are attributed to errors in the calibration of the qubit measurement setup and to fluctuations of the qubit cavity resonance frequency. Additionally, a quantum process is reconstructed via a maximum-likelihood fit, leading to an overall mean state fidelity of $\overline{\mathcal{F}}_{\text{s,}\chi}=(96.3\pm0.6)\,\%$.

Another important property of a NPQD is to preserve the temporal waveform of the photonic qubit (Fig.\,\ref{fig:FOM2}d). This is of great importance in scenarios where photons interfere or where quantum information is encoded in the temporal mode of the photon in addition or instead of the polarization. In order to illustrate that our NPQD preserves the waveform, we compare three different envelopes of qubit pulses reflected off our atom-cavity system (blue points) with pulses reflected off an empty cavity (yellow points). For all three cases we find an intensity waveform overlap between the outgoing and the incoming photon exceeding $99.5\,\%$.

\begin{figure}
\centering
\includegraphics[width=0.5\linewidth]{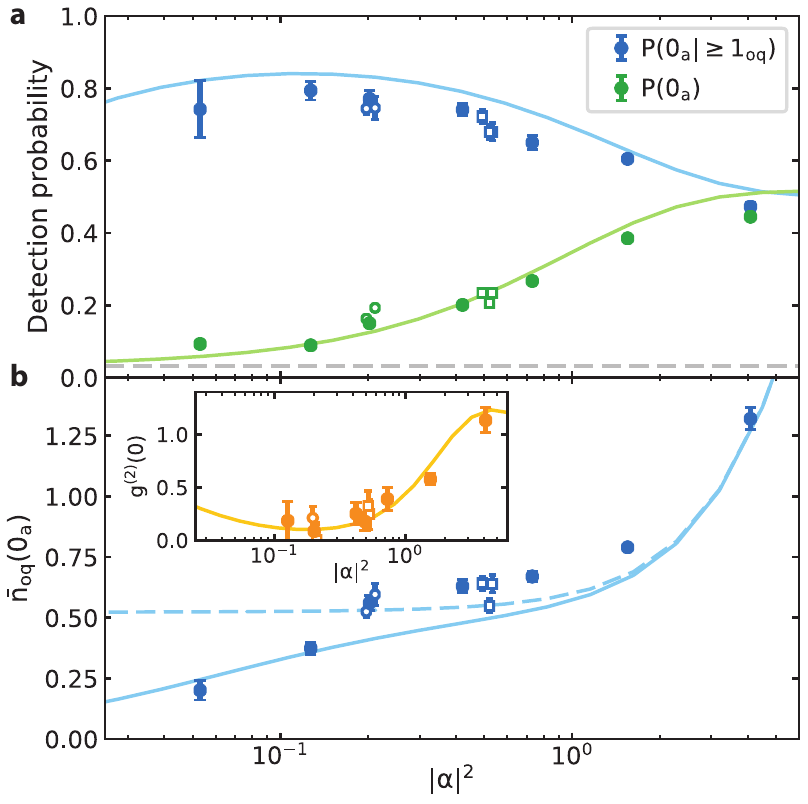}
 \caption{\textbf{Characterization with different mean input photon numbers.} \textbf{a}, Nondestructive detection probability as a function of the mean input photon number $\left|\alpha\right|^2$ in front of the qubit cavity. $P(0_\text{a}|\geq1_\text{oq})$ is conditioned on a successful qubit reflection ($\geq1_\text{oq}$) while $P(0_\text{a}$) is unconditioned. The horizontal dashed line represents the NPQD dark-count probability $p_\text{DC}$. \textbf{b}, Mean photon number at the qubit cavity output conditioned on a nondestructive detection event $\bar{n}_\text{oq}(0_\text{a})$. The inset pictures the autocorrelation function $g^{(2)}(\tau=0)$ of the nondestructively detected qubits. All solid data points are taken with Gaussian photon pulse shapes and linear near-vertical polarization. Open circle (square) data points were taken with orthogonal polarization (different pulse shapes, Fig.\,\ref{fig:FOM2}c). For all plots the solid line is given by theoretical simulations described in Supplementary Information. The dashed line in \textbf{b} considers $p_\text{DC}=0$. The error bars in \textbf{a} represent the $1\sigma$ confidence interval. The error bars in \textbf{b} (inset) represent the standard error (standard deviation).} 
\label{fig:FOM} 
\end{figure}

The mean input photon number $\left|\alpha\right|^2$ of the weak coherent pulses encoding the qubit plays a role for the performance of the NPQD. Its characterization is presented in Fig.\,\ref{fig:FOM} where different figures of merit of our detector are shown.
For the given range of $\left|\alpha\right|^2$ in front of the qubit cavity, the nondestructive detection probability upon qubit survival $P(0_\text{a}|\geq1_\text{oq})$ (blue points in Fig.\,\ref{fig:FOM}a) shows a maximum of $(79\pm3)\,\%$ at $\left|\alpha\right|^2=0.13$. Here, the major error is attributed to a difference in intensity reflection coefficients for a coupling ($R_{\left|0_\text{a}\right>}=0.50\pm0.02$) and a noncoupling atom ($R_{\left|1_\text{a}\right>}= 0.117\pm0.003$), see Methods. For high $\left|\alpha\right|^2$, the conditional probability converges towards $0.5$ due to the balanced contribution of odd and even photon numbers \cite{Daiss2019}, and decreases for small $\left|\alpha\right|^2$ values due to dark counts of the qubit measurement setup. The unconditional probability $P(0_\text{a})$ (green points) shows the same convergence for high $\left|\alpha\right|^2$. However, in the limit of $\left|\alpha\right|^2\rightarrow0$, the probability is lower-bounded by $p_\text{DC}$ (grey dashed line). The latter is due to imperfections in the atomic state manipulation, in the optical pumping and in the state detection.

Another important figure of merit is the probability to have a photonic qubit at the NPQD output conditioned on its nondestructive detection, $P(1_{\rm{oq}}|0_\text{a})$. Since we characterize the NPQD with weak coherent states (and not with single-photon Fock states), we instead show in Fig.\,\ref{fig:FOM}b the conditioned mean photon number, $\bar{n}_\text{oq}\left(0_\text{a}\right)$, which is equivalent to $P(1_{\rm{oq}}|0_\text{a})$ for small input photon numbers.
We observe $\bar{n}_\text{oq}\left(0_\text{a}\right)=0.56\pm0.02$ for $\left|\alpha\right|^2=0.2$, but this value decreases for smaller $\left|\alpha\right|^2$ due to the NPQD dark counts $p_\text{DC}$. The simulation that considers $p_\text{DC}=0$ (dashed line), converges towards $P(1_{\rm{oq}}|0_a)=52.3\,\%$ for $\left|\alpha\right|^2\rightarrow 0$. This value is below one because of parasitic losses of the cavity mirrors, imperfect mode matching and a finite atomic decay rate.
Interestingly, not only the mean photon number but also the photon statistics changes after qubit reflection (inset in Fig\,\ref{fig:FOM}b). We prove this by measuring the autocorrelation function of the reflected qubits conditioned on their nondestructive detection. The obtained sub-Poissonian statistics $g^{(2)}(0)<1$ originates from the distillation of single photons out of the incoming weak coherent pulse (see Supplementary Information and \cite{Daiss2019}).

\begin{figure}
\centering
\includegraphics[width=0.5\linewidth]{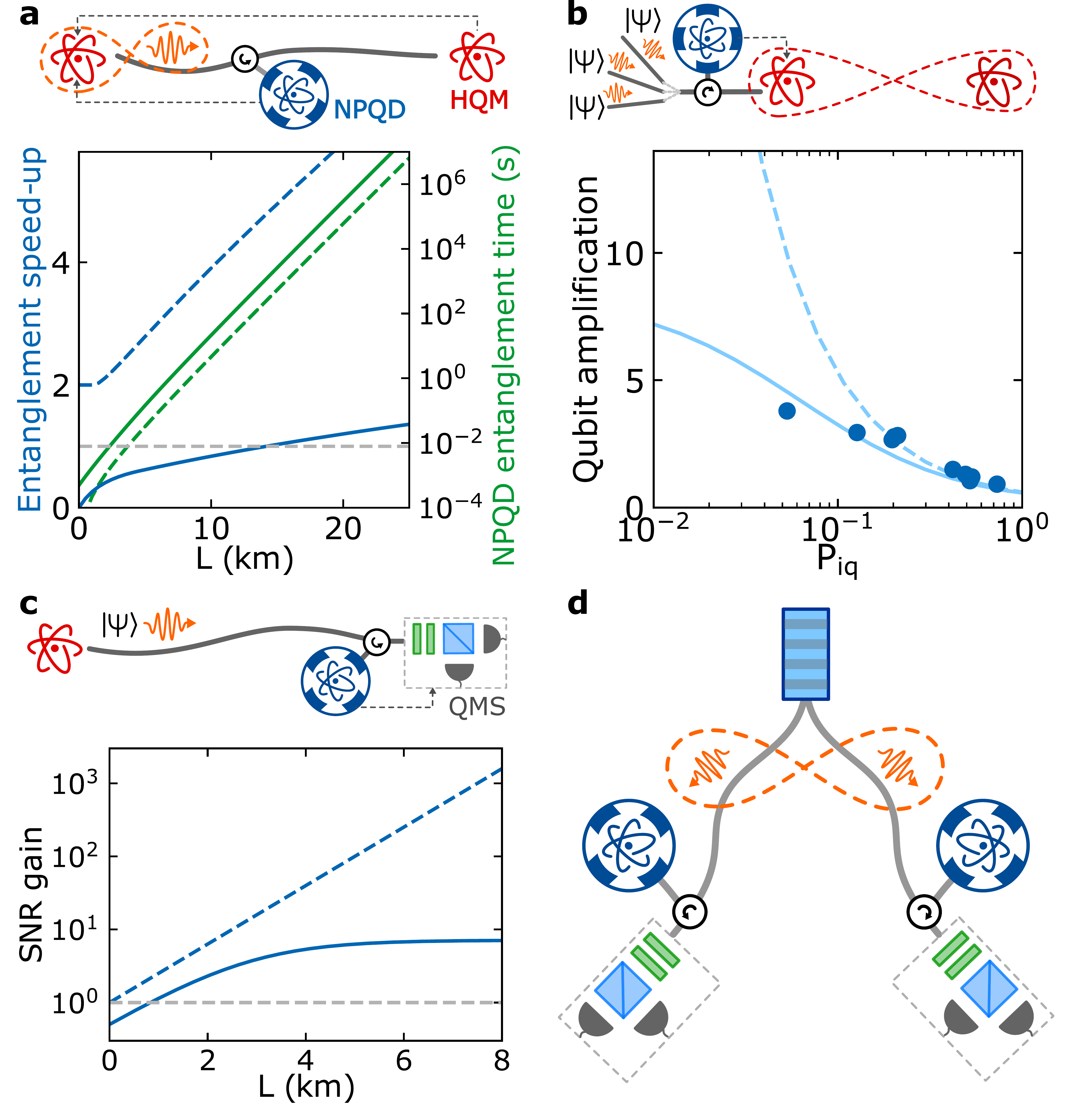}
\caption{\textbf{NPQD applications.} \textbf{a}, Entanglement generation between an atom-photon entanglement source and a heralded quantum memory (HQM). A NPQD along the transmission channel can speed up the mean sender-receiver entanglement time compared to the standard situation. \textbf{b}, Photonic qubits are remotely sent to a receiver for a follow-up operation that involves precious resources (e.g. long distance entanglement). The qubit amplification gives the ratio of probabilities to have a photon before $P_\text{iq}$ and after its nondestructive detection at the NPQD. \textbf{c}, Photonic qubits transfer to a noisy qubit measurement setup (QMS). Gating the QMS with the NPQD herald signal improves the signal-to-noise ratio (SNR). \textbf{d}, Bell test based on precertification of the photon’s presence \cite{Sciarrino2012}. Dashed lines in plots represent the situation of a perfect NPQD (in \textbf{a} and \textbf{c}) or a dark-count-free NPQD (in \textbf{b}).} 
\label{fig:applications}
\end{figure}

To explore the potential of our NPQD, we exemplary discuss now four applications that would benefit from our detector (details in Methods). Fig.\,\ref{fig:applications}a and b illustrate situations in which monitoring the qubit loss along a transmission channel allows saving time and precious resources.
The first example (Fig.\,\ref{fig:applications}a) consists of an atom-photon entanglement source \cite{Rempe2015} that sends photonic qubits to a heralded quantum memory \cite{Brekenfeld2020} in order to generate sender-receiver entanglement. The plot shows the entanglement speed-up defined as the ratio between the mean entanglement generation time without and with the nondestructive detector, $T_\text{ent}/T^\text{NPQD}_\text{ent}$, versus the channel length $L$. The position of the detector was chosen such that $T^\text{NPQD}_\text{ent}$ is minimal. Our nondestructive detector (solid lines) outperforms direct transmission at channel distances $\geq14\,$km whereas a perfect NPQD (dashed lines) would provide an advantage at any distance. 
The second example (Fig.\,\ref{fig:applications}b) describes a situation in which photonic quantum states are remotely sent to a receiver that wants to perform operations on them using precious resources (e.g. long-distance entanglement). A NPQD right before the receiver allows performing the operation only when the photonic qubit has not been lost along the way. Here, a good figure of merit is the ratio of the probability to have a reflected qubit photon after nondestructive detection $P(1_\text{oq}|0_\text{a})$ and the probability of having an incoming qubit photon $P_\text{iq}$, usually called qubit amplification \cite{Pryde2013}. The qubit amplification is significantly higher than one for $P_\text{iq}\ll1$ and would notably improve in the absence of NPQD dark counts (dashed line).

Another group of applications refer to situations in which NPQDs can improve a subsequent photonic qubit measurement. Fig.\,\ref{fig:applications}c shows a scenario in which photonic qubits are sent to a remote receiver that uses noisy detectors for the qubit measurement. A NPQD right before the receiver allows measuring the qubit only when it was not lost, reducing the impact of measurement noise. The signal-to-noise ratio (SNR) gain is defined as the ratio of the SNR with and without a NPQD, SNR$_\text{NPQD}$/SNR, and exceeds one for a sender-receiver distance $L>1\,$km. For longer distances, this ratio converges towards approximately seven which could be raised with lower NPQD dark counts (illustrated by the dashed line which considers perfect NPQD parameters). A last example relates to loophole-free Bell tests (Fig.\,\ref{fig:applications}d). As proposed in \cite{Sciarrino2012}, the herald signals of NPQDs could allow two parties to be certain that they share an entangled photon pair right before the measurement, helping to close the detection loophole. The important parameter here is the photonic qubit reflection probability conditioned on its nondestructive detection. From our experiment we find $\bar{n}_\text{oq}(0_\text{a})=0.56\pm0.02$ for an input $\left|\alpha\right|^2=0.2$ (Fig.\,\ref{fig:FOM}b). This value exceeds the minimum detection efficiency of $43\,\%$ (assuming no detector background noise) which is required for a detection-loophole-free asymmetric Bell test \cite{Brunner2007}.

In conclusion, we have demonstrated a nondestructive detector for photonic polarization qubits. We anticipate that it should also work with time-bin qubits. The main features are a conditional detection efficiency of up to $(79\pm3)\,\%$ and a qubit preservation fidelity of at least $(96.2\pm0.3)\,\%$. None of the observed limitations seems fundamental. Most important, we presented four possible applications that would already benefit from our present still imperfect device. This creates confidence that our detector will turn out to be a useful tool in near-future quantum communication links and fundamental tests of quantum physics.


\clearpage
\printbibliography
\newpage

\clearpage
\appendix
\section*{Methods}
\begin{refsection} 
\section{NPQD experimental sequence}
The experiment starts with a two seconds long loading phase of a single atom into the crossing point of the two fibre-cavity modes (details in \cite{Brekenfeld2020}). A magneto-optical trap with $^{87}$Rb atoms is loaded $\approx10\,$mm above the fibres and is released such that the laser-cooled atoms fall towards the intracavity region. At the crossing point of the two cavity modes there is a three dimensional optical lattice where single atoms are trapped. The lattice consists of two blue detuned intracavity traps  ($774.6\,$nm and $776.5\,$nm wavelength) and one red detuned standing-wave optical dipole trap ($799.2\,$nm wavelength) with respect to the $D$-line and with a trap depth of $U_0/k_\text{B}\approx 1\,$mK each. A cooling beam is coming in under an oblique angle and cools the single atoms in the trap, where they remain a few seconds. Cooling light that is scattered by the atom into the state-detection cavity mode, is subsequently detected with a single photon detector so that the presence of single atoms is confirmed. Afterwards, the NPQD scheme runs with a repetition rate of $576\,$Hz. This sequence divides into three parts: the atomic cooling, atomic state preparation and the NPQD part. The temporal length of atomic cooling constitutes $97\,\%$ of the fast sequence time. This high fraction goes along with a desired and small microwave (MW) duty cycle to minimize system heating and cavity shaking that is attributed to the applied MW fields. However, the relatively long cooling time is used to intermediately conduct six times a $3\,\mu$s long sequence to generate single photons emitted by the atom. The recorded photon counts are used to calculate the autocorrelation function $g^{(2)}(0)=1-1/n$ in order to infer the number of trapped atoms $n$. The atom cooling is followed by a $30\,\mu$s long optical pumping phase to prepare the atom in Zeeman state \ket{F=1, m_F=0}. Here, the applied laser fields are the same as described in \cite{Brekenfeld2020}, except for the finally applied $4\,\mu$s long $\pi$ polarized laser field mentioned in it. After optical pumping, the sequence starts with the NPQD scheme as reported in the main text. We have stated that for cavity-assisted state detection the state-detection cavity is near resonant with the atomic transition $D_2$: \ket{F=2} $\leftrightarrow$ \ket{F'=3}. Since the state-detection cavity exhibits an intentional polarization mode splitting by more than four cavity linewidths \cite{Brekenfeld2020}, only the $\pi$ polarization mode is near resonant with the atomic transition whereas the second polarization mode is blue detuned.

\section{Coherent atomic state manipulation}
The nondestructive detector scheme requires the coherent manipulation of the atomic ground states \ket{F=1, m_F=0} and \ket{F=2, m_F=0}. In our experiment this manipulation is done via microwave radiation. It is advantageous to use microwave (MW) fields as opposed to a Raman transition with near infrared radiation fields as the latter is intrinsically less detuned from the excited states and therefor leads to a higher probability to populate these which causes decoherence processes \cite{Koerber2018}.

\begin{figure}[h!]
\centering
\includegraphics[width=0.5\linewidth]{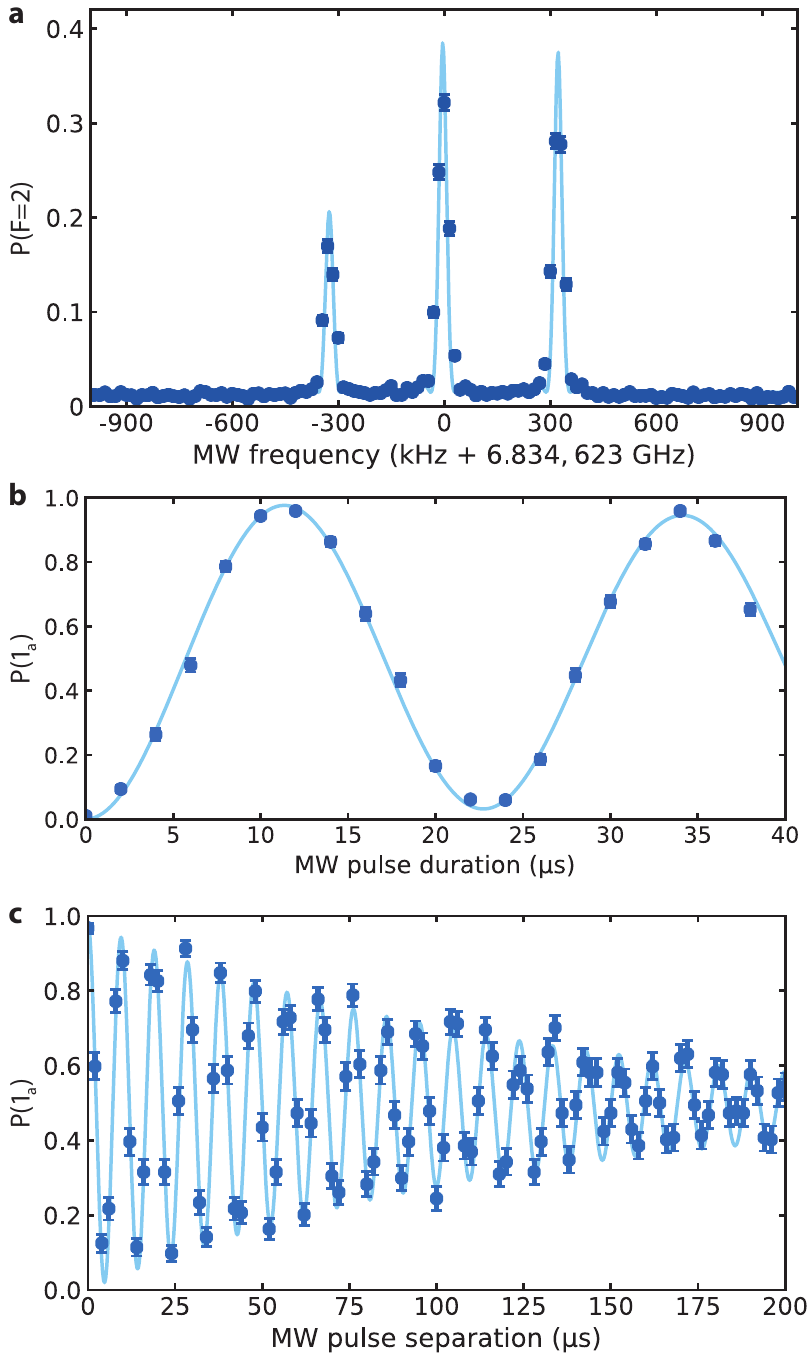}
\caption{\textbf{Coherent microwave driving.} \textbf{a}, Microwave (MW) spectroscopy of an atom prepared in Zeeman state \ket{F=1, m_F=0}. The light blue curve is a fit consisting of three $\sinc^2$ functions. Fit parameters are center and Rabi frequency of each transition whereas the pulse duration is fixed to $25\,\mu$s. The peak at $0$\,kHz corresponds to the \ket{0_\text{a}} $\leftrightarrow$ \ket{1_\text{a}} transition. \textbf{b}, MW-driven Rabi oscillation on the \ket{0_\text{a}} $\leftrightarrow$ \ket{1_\text{a}} transition for an atom prepared in \ket{0_\text{a}}. \textbf{c}, An atom prepared in state \ket{0_\text{a}} is subsequently driven into a superposition of \ket{0_\text{a}} and \ket{1_\text{a}} by a MW $\pi/2$ pulse. After a variable time a second MW $\pi/2$ pulse is applied. During the waiting time the internal MW clock is shifted by $100$\,kHz. Final state detection measures the population in \ket{1_\text{a}}. All error bars represent the $1\sigma$ confidence interval.}
\label{fig:MW}
\end{figure}

We have a MW antenna inside the vacuum chamber with a distance of $<2\,$cm  to the fibre-cavity modes crossing point. The antenna consists of a single loop with a circumference of $4.4\,$cm equal to the MW field wavelength. A magnetic guiding field of $226.5\,$mG magnitude defines the quantization axis along the qubit cavity axis which allows the addressing of a single ground state transition at a MW pulse duration of $5.8\,\mu$s. Fig.\,\ref{fig:MW}a shows a MW spectrum for an atom that is prepared in state \ket{F=1, m_F=0} and which was subsequently driven with a MW field of varying frequency and a rectangular pulse duration of $25\,\mu$s. Final cavity-assisted state detection measures the population in state $5^2S_{1/2}$ $F=2$.

Choosing the related frequency enables the driving of the transition of interest: \ket{F=1, m_F=0} $\leftrightarrow$ \ket{F=2, m_F=0}. The coherent Rabi flopping of these two states is shown in Fig.\,\ref{fig:MW}b, revealing the $\pi/2$ pulse duration of $5.8\,\mu$s. One limitation that affects this coherent driving is decoherence, which leads to a shrinking of the Rabi oscillations visibility. To further characterize this effect, we performed a Ramsey type experiment (Fig.\,\ref{fig:MW}c). We have applied two rectangularly shaped $\pi/2$ pulses with varying time gap in between. During the waiting time, the internal MW clock is shifted by $100\,$kHz to distinguish the decoherence from small MW detunings. Decoherence is then unambiguously detected by the shrinking visibility of the oscillations. Fig.\,\ref{fig:MW}c shows how the visibility decreases to the $1/e$ value after $121\,\mu$s which we define as the coherence time. We suspect the decoherence coming from effects of a mechanical non-ground state cooled atom that is trapped via a near resonant red-detuned optical dipole trap. Hence, the hyperfine ground states \ket{F=1} and \ket{F=2} are differently AC Stark shifted which leads to differential trap frequencies and ultimately to non-degenerate ground state transition frequencies. This situation can be improved by ground state cooling of the single atom \cite{Reiserer2013_2} or a further red-detuned (and therefor a more powerful) optical dipole trap.

\section{Atom-cavity interaction and conditional reflection}
\label{sec:condRefl}
\begin{figure}
\centering
\includegraphics[width=0.75\linewidth]{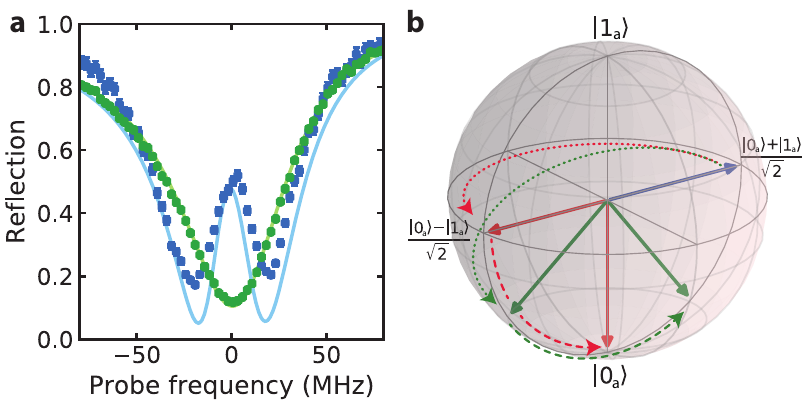}
\caption{\textbf{Atom-cavity reflection spectra and the atomic state during the NPQD scheme.} \textbf{a}, Cavity reflection spectra of the qubit cavity with no atom (green points) and with a cavity mode that is strongly coupled to an atom prepared in \ket{F=1, m_F=0} (blue points). The solid lines represent fit functions and the error bars show the standard deviation. \textbf{b}, Bloch sphere with the atomic state (represented by vectors) at different stages of the detector scheme. The blue vector is the initial atomic state after the first $\pi/2$ microwave pulse. After photon reflection, the initial state turns into the red or green state (dotted arrow as a guide to the eye). Green (red) states show the situation with(out) conditional reflection. A subsequent MW $\pi/2$ pulse rotates the atom to the final state (dashed arrow follows the rotation). State detection (not shown) then projects the atomic state onto \ket{0_\text{a}} or \ket{1_\text{a}}.}
\label{fig:condRefl}
\end{figure}

One enabling ingredient for the NPQD is the polarization-independent strong coupling between the atom and the photonic qubit mediated by the qubit cavity. For characterization, reflection spectra were measured (Fig.\,\ref{fig:condRefl}a) with a probe field in a superposition of left and right circular polarization, once with an atom coupled to the cavity mode (blue points) and once without any atom (green points). In the figure, a probe field frequency of zero corresponds to the atomic transition frequency. The empty cavity spectrum yields a field decay rate of $\kappa_\text{QC}=2\pi\times\left(34.6\pm0.3\right)$\,MHz at a qubit cavity length of $162\,\mu$m. The normal-mode spectrum provides a coupling rate of $g=2\pi\times\left(18.6\pm0.5\right)$\,MHz which leads to, considering the atomic dipole decay rate of $\gamma =2\pi\times3$\,MHz, a cooperativity of $C=g^2/(2\kappa_\text{QC}\gamma)=1.67\pm0.09$. Note the difference between the intensity reflection coefficients with ($R_{\left|0_\text{a}\right>}=0.50\pm0.02$) and without ($R_{\left|1_\text{a}\right>}= 0.117\pm0.003$) coupling atom, both at zero detuning. The latter constitutes a significant error source of our nondestructive detector which leads to a lower detection probability which is explained in the following. 

The theoretical model given in Supplementary Information includes a detailed set of contributing imperfections of our NPQD. However, here we assume a perfect system detecting single photons with only the conditional reflection as imperfection to clarify its consequences for the detector. The NPQD scheme starts with the atom prepared in a superposition of two ground states $\left|\Psi_\text{a1}\right>=(\left|0_\text{a}\right>+\left|1_\text{a}\right>)/\sqrt{2}$ which is ideally turning into $\left|\Psi_\text{a2}\right>=(\left|0_\text{a}\right>-\left|1_\text{a}\right>)/\sqrt{2}$ upon successful photon reflection (blue and first red vector in Fig.\,\ref{fig:condRefl}b). However, due to the different cavity reflection coefficients corresponding to the two atomic states, the initial state turns into
\begin{equation}
\left|\Psi_{\text{a}2}\right>=\frac{r_0\left|0_\text{a}\right>+r_1\left|1_\text{a}\right>}{\sqrt{\left|r_0\right|^2+\left|r_1\right|^2}}
\end{equation}
with $r_0=\left|\sqrt{R_{\left|0_\text{a}\right>}}\right|$ and $r_1=\left|\sqrt{R_{\left|1_\text{a}\right>}}\right|e^{i\pi}$ (first green vector). As $r_0$ and $r_1$ are not only different in phase but also in magnitude, the atomic state leaves the equatorial plane of the Bloch sphere. A subsequent MW $\pi/2$ pulse rotates this state into $\left|\Psi_{\text{a}3}\right>=\hat{R}_a(\pi/2)\left|\Psi_{\text{a}2}\right>$ which goes beyond the pole \ket{0_\text{a}} (second green vector) and therefor leads to a state that is not orthogonal to \ket{1_\text{a}}. The detection probability upon qubit reflection is then upper-bounded by $P(0_\text{a}|1_\text{oq})=\left|\left<0_\text{a}|\Psi_{\text{a}3}\right>\right|^2=89\,\%$. The further reduction of this value that we observe is attributed to a group of imperfections that are further discussed in Supplementary Information.

\section{Photonic qubit fidelity and cavity birefringence}
\begin{figure}
\centering
\includegraphics[width=0.75\linewidth]{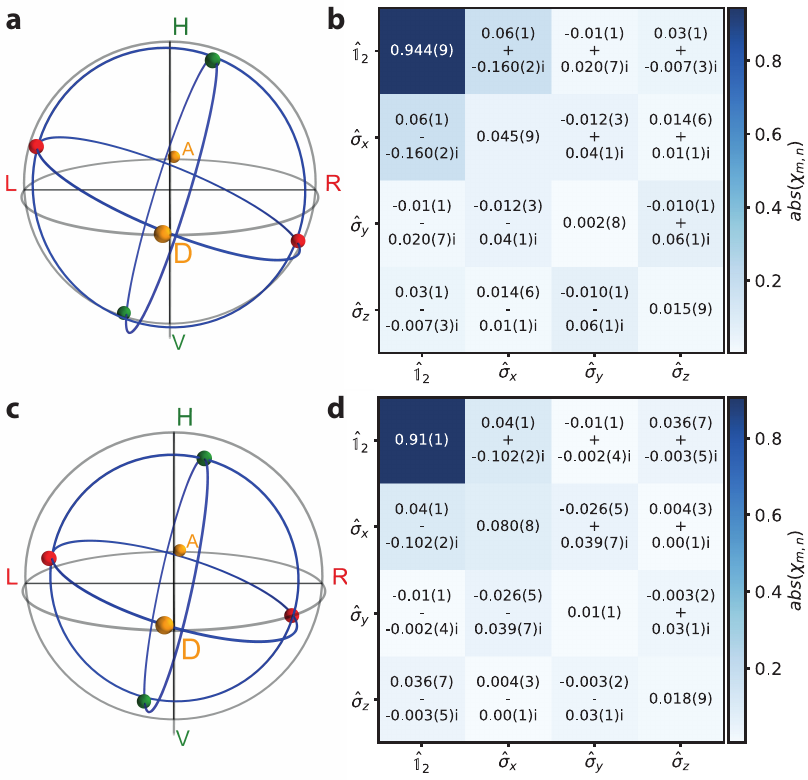}
\caption{\textbf{Polarization state and quantum process tomography.} \textbf{a} and \textbf{b} consider photon counts at the tomography setup which are conditioned on the nondestructive qubit detection. For comparison, \textbf{c} and \textbf{d} do not condition on the nondestructive detection.
\textbf{a} (\textbf{c}), Poincar\'{e} sphere showing the underlying quantum process of our NPQD. The coloured spheres are the result of polarization state tomography for reflected photonic pulses for the correspondingly coloured and labelled input polarizations. \textbf{b} (\textbf{d}), NPQD underlying quantum process matrix $\chi$ which is reconstructed via a maximum-likelihood fit. The rotation around the $A/D$ axis is described by an operator built by the Pauli matrix $\sigma_x$. The uncertainty of $\chi_{m,n}$ is assessed with a Monte Carlo method and represents the standard error.}
\label{fig:QPTmatrix}
\end{figure}

As shown in Fig.\,\ref{fig:FOM2}, the photonic polarization qubit experiences a small rotation after its nondestructive detection. This is due to the birefringence of the qubit cavity, which originates from a small ellipticity of the fibre-cavity mirrors and leads to a polarization mode splitting \cite{Uphoff2015} of a fifth of the qubit cavity linewidth. By means of a $\lambda/2$ retardation waveplate positioned after the cavity, all polarization states rotate such that the qubit cavity eigenmode polarizations at reflection rotate into polarizations $A$ and $D$ at the detection setup. We measure that a superposition of polarizations $A$ and $D$ experiences a rotation of $42\,^\circ$ around the $A/D$ axis on the Bloch sphere in case of an empty cavity. Since the detector scheme relies on an atom being in a superposition of a coupling and noncoupling state, the incoming photonic qubit does not experience this full rotation. A rotation angle of $19.6\,^\circ$ is extracted from a fit of the input and output polarization states which are measured by polarisation state tomography (Fig.\,\ref{fig:QPTmatrix}a, the analysis follows the description of \cite{Brekenfeld2020}).
This rotation is also observed in Fig.\,\ref{fig:QPTmatrix}b which shows the matrix of the related detector underlying quantum process \cite{Nielsen2010}. The latter is derived from a maximum-likelihood fit with matrix elements uncertainties that are calculated according to \cite{Brekenfeld2020}. Major contributing matrix elements are $\chi_{0,0}$, $\chi_{1,0}$ and $\chi_{0,1}$, illustrating the polarization rotation effect. 

In the main text we provide the mean state fidelities under the condition of nondestructive detection. However, evaluation of unconditional SPD counts (e.g. events from all photonic qubits reflected from the NPQD) reveals a reduction of the polarization state fidelity for cavity noneigenstate polarizations (compare $\mathcal{F}_\text{cond.}$ with $\mathcal{F}_\text{uncond.}$ in Table\,\ref{tab:qubitfidelities}). We attribute this to a partial entanglement effect between the photonic polarization state and the atomic state since the polarization rotation preferably occurs when the atom is in the noncoupling state. This effect can be theoretically described as follows. An initial atomic superposition state and an incoming photonic qubit lead to a partially entangled state after reflection,
\begin{equation}
\left|\Psi\right>=\frac{r_0\left|\Psi_\text{ph}\right>\left|0_\text{a}\right>+r_1(\hat{R}\otimes\hat{\mathbb{1}})\left|\Psi_\text{ph}\right>\left|1_\text{a}\right>}{\sqrt{|r_0|^2+|r_1|^2}}.
\label{eq:entangledstate}
\end{equation}
\ket{0_\text{a}} (\ket{1_\text{a}}) represent the (non)coupling atomic state and \ket{\Psi_\text{ph}} represents the photon polarization state, which experiences a rotation $\hat{R}$ when the atom is in state \ket{1_\text{a}}. $r_0$ and $r_1$ describe the field reflection coefficients as used in section\,\ref{sec:condRefl}. Moreover, we assume no qubit rotation in case of a coupling atom.
When the atomic part is not observed, the entanglement translates into decoherence for the photonic qubit as shown by the shrinking sphere in Fig.\,\ref{fig:QPTmatrix}c. This situation holds only true for polarization states that experience a rotation due to the cavity birefringence. Since photonic polarizations $A$ and $D$ do not rotate, for these polarization the state in eq.\,\ref{eq:entangledstate} becomes separable and therefor follows the ideal situation.

As mentioned before, the polarization rotation leads to a qubit infidelity that could be overcome by means of retardation waveplates. However, in addition one would also expect a polarization dependency on the imprinted atomic phase shift at qubit reflection which would affect the nondestructive detection efficiency. We have tried to observe this effect by comparing the detector performance for different incoming polarization states (Fig.\,\ref{fig:FOM}), but can not find significant differences in the NPQD performance. We attribute this observation to the fact that the birefringence effect is relatively small compared to other imperfections, e.g. the conditional reflection.

\section{Theory models for NPQD applications}
\label{sec:theoryApplications}
In Fig.\,\ref{fig:applications} we show simulations of specific situations in which our nondestructive detector is advantageous. The details of these simulations are explained in this section. As the operating wavelength of our NPQD is $780\,$nm, we consider photonic qubits of this wavelength with a corresponding fibre attenuation of $\alpha=4$\,dB/km. 

Situation 1 (Fig.\,\ref{fig:applications}a), long distance entanglement between an atom-photon entanglement source and an heralded quantum memory: For simplicity we consider the ideal situation in which the time required to entangle the two systems is given by the communication time (e.g. the time to distribute the entangled photonic qubit plus the time to communicate back if the heralded storage succeeded). In such a situation the mean entanglement time is given by $T_\text{ent} = 2L/(c\cdot p_\text{ent})$, where $L$ is the distance between sender and receiver, $c$ is the speed of light in an optical fibre and $p_\text{ent}=\eta_\text{AP}10^{-\alpha L/10}\eta_\text{H}$ is the heralded entanglement distribution probability. This probability depends on the atom-photon entanglement source efficiency $\eta_\text{AP}$, the attenuation coefficient of the transmission channel $\alpha$, and the heralding efficiency of the heralded quantum memory $\eta_\text{H}$. When a NPQD is inserted along the transmission channel (at a distance $l$ from the sender), the communication distance is shortened in the cases where the photonic qubit is lost before the NPQD. One can express the average communication distance as a function of the nondestructive qubit detection probability $P(0_\text{a})$ as $\left\langle l \right\rangle=P(0_\text{a})L+[1-P(0_\text{a})](l+t_{\rm{NPQD}}\cdot c/2)$. This expression  will replace the total communication distance $L$ used in the expression of $T_\text{ent}$ in order to obtain the heralded entanglement distribution time using a NPQD ($T_\text{ent}^{\rm{NPQD}}$). $t_{\rm{NPQD}}$ is the time the NPQD requires to perform the nondestructive detection operation, which delays the subsequent classical communication. The probability for a nondestructive detection event $P(0_\text{a})$ is given by 
\begin{align}
P(0_\text{a}) = \eta_\text{AP} 10^{-\alpha l/10} P(0_\text{a}|1_\text{iq})\cdot(1-p_\text{DC})+p_\text{DC}
\end{align}
which takes into account the nondestructive detection probability upon an incoming qubit $P(0_\text{a}|1_\text{iq})$ and the NPQD dark counts $p_\text{DC}$. In addition, the entanglement probability is also reduced by the inefficiency of the nondestructive detector, $p_\text{ent}^\text{NPQD}=p_\text{ent}\cdot P(0_\text{a}|1_{\rm{iq}})\cdot P(1_{\rm{oq}}|0_\text{a})$. With these expressions, we calculate the average time to entangle the two systems when using a NPQD, $T_\text{ent}^{\text{NPQD}} = 2\left\langle l \right\rangle/(c\cdot p_\text{ent}^{\text{NPQD}})$, and compare it with the situation that does not include a nondestructive detector $T_\text{ent}$, in order to obtain the entanglement speed-up. For both simulations we consider $\eta_{\rm{AP}}=0.5$ and $\eta_\text{H}=0.11$, which are realistic parameters that have been obtained in \cite{Ritter2012} and \cite{Brekenfeld2020}.

Situation 2 (Fig.\,\ref{fig:applications}b), qubit amplification for different input photon number probabilities: The qubit amplification $A_\text{q}$ is defined as the ratio of the probability to have a photonic qubit at the output of the NPQD conditioned on its nondestructive detection $P(1_\text{oq}|0_\text{a})$ and the probability to have an input qubit  $P_\text{iq}$. These quantities are related to the ones shown in Fig.\,\ref{fig:FOM}, from which the qubit amplification is inferred as $A_\text{q}=\bar{n}_\text{oq}(0_\text{a})/|\alpha|^2$ for small $|\alpha|^2$. The data points shown in Fig.\,\ref{fig:applications}b are calculated according to this expression. In order to simulate the qubit amplification with our NPQD parameters (solid line) and without NPQD dark counts (dashed line), we make use of the theory model described in Supplementary Information.
 
Situation 3 (Fig.\,\ref{fig:applications}c), remote photonic qubits are measured with noisy detectors: a NPQD gates a qubit measurement in order to improve its signal-to-noise ratio $\text{SNR}=p_\text{s}/p_\text{n}$. This gating reduces the noise detection probability to $p_\text{n}^\text{NPQD}=P(0_\text{a})\cdot p_\text{n}$ but also reduces the signal due to the parasitic losses induced by the nondestructive detector $p_\text{s}^{\rm{NPQD}}=p_\text{s}\cdot  P(0_\text{a}|1_{\rm{iq}})\cdot P(1_{\rm{oq}}|0_\text{a})$. Additionally, we relate the nondestructive detection probability with our NPQD parameters, $P(0_\text{a})=P(0_\text{a}|1_\text{iq})\cdot p_\text{s}\cdot (1 - p_{\rm{DC}}) + p_{\rm{DC}}$, and consider that the input signal probability depends only on the transmission losses of the communication channel, $p_\text{s}=10^{-\alpha L/10}$. This allows calculating the SNR gain as a function of the communication distance $L$, considering a NPQD with our parameters (solid line) and a perfect NPQD (dashed line).

\section*{Data availability}
The data that support the findings of this study are available from the corresponding author upon reasonable request.

\printbibliography[heading=subbibliography, title={Methods references}]
\end{refsection}

\subsection*{Acknowledgements}
We thank M. Brekenfeld and J. D. Christesen for contributions during an early stage of this work. This work was supported by the Bundesministerium für Bildung und Forschung via the Verbund Q.Link.X (grant no. 16KIS0870), the Deutsche Forschungsgemeinschaft under Germany’s Excellence Strategy (EXC-2111, 390814868) and the European Union’s Horizon 2020 research and innovation programme via the project Quantum Internet Alliance (GA no. 820445). P. F. acknowledges support by the Cellex-ICFO-MPQ postdoctoral fellowship program.

\subsection*{Author contribution}
All authors contributed to the experiment, analysis of the results and writing of the manuscript.

\subsection*{Competing interests}
The authors declare no competing interests.

\subsection*{Additional information}
Supplementary Information text is available for this paper.\newline
Correspondence and requests for materials should be addressed to \\D. N. (dominik.niemietz@mpq.mpg.de).

\newpage
\section*{Extended Data}

\clearpage
\begin{singlespace}
\renewcommand{\arraystretch}{1.5}
\begin{table}
\caption{Photonic polarization qubit fidelities}
\label{tab:qubitfidelities}
\begin{threeparttable}
\renewcommand{\TPTminimum}{\linewidth}
\makebox[\linewidth]{%
\begin{tabular}{ccc} \toprule
    Input & $\mathcal{F}_\text{cond.}$\tnote{*} & $\mathcal{F}_\text{uncond.}$\tnote{$\dagger$}\\ \midrule
    \ket{H}  & $0.94\pm0.01$  & $0.92\pm0.01$ \\
    \ket{V}  & $0.969\pm0.008$  & $0.92\pm0.01$  \\
    \ket{A}  & $0.985\pm0.004$  & $0.985\pm0.004$  \\
    \ket{D}  & $0.986\pm0.004$  & $0.978\pm0.004$ \\
    \ket{R}  & $0.944\pm0.009$  & $0.90\pm0.01$ \\
    \ket{L}  & $0.943\pm0.009$  & $0.910\pm0.009$ \\ \bottomrule
    $\overline{\mathcal{F}}_\text{all}$\tnote{$\ddagger$} & $0.962\pm0.003$ & $0.936\pm0.003$\\
    $\overline{\mathcal{F}}_{\circlearrowleft}$\tnote{$\mathsection$} & $0.980\pm0.003$ & $0.944\pm0.003$\\ \midrule
    $\mathcal{F}_\chi$\tnote{$\mathparagraph$} & $0.944\pm0.009$ & $0.91\pm0.01$\\
    $\overline{\mathcal{F}}_{\text{s,}\chi}$\tnote{**} & $0.963\pm0.006$ & $0.938\pm0.007$\\ \bottomrule
\end{tabular}}
\begin{tablenotes}
\small
\item[*] SPD counts are conditioned on the nondestructive qubit detection
\item[$\dagger$] SPD counts are unconditionally evaluated
\item[$\ddagger$] Mean value of all six input polarization states
\item[$\mathsection$] An inverse unitary rotation around the $A/D$ axis is applied to the measured output states before calculating the mean fidelity
\item[$\mathparagraph$] Quantum process fidelity
\item[**] Mean state fidelity inferred from the quantum process fidelity via
$\overline{\mathcal{F}}_{\text{s,}\chi}=\frac{2\mathcal{F}_\chi+1}{3}$
\end{tablenotes}
\small
Statistical errors : The polarization state fidelities, $\overline{\mathcal{F}}_\text{all}$ and $\overline{\mathcal{F}}_{\circlearrowleft}$ are assigned with a $1\sigma$ confidence level accounting for statistical uncertainties due to the finite number of detected photons. The uncertainty of $\mathcal{F}_\chi$ is assessed with a Monte Carlo method and represents the standard error. 
\end{threeparttable}
\end{table}
\end{singlespace}

\end{document}


\title{Supplementary Information to: Nondestructive detection of photonic qubits}
\author[1]{D. Niemietz}
\author[1,2]{P. Farrera}
\author[1]{S. Langenfeld}
\author[1]{G. Rempe}
\affil[1]{Max Planck Institute of Quantum Optics, Hans-Kopfermann-Str. 1, 85748 Garching, Germany}
\affil[2]{ICFO-Institut de Ciencies Fotoniques, The Barcelona Institute of Science and Technology, 08860 Castelldefels (Barcelona), Spain}
\date{\today}
\maketitle

\section{Theory model for the NPQD}
\label{sec:theoryNPQD}
The measured parameters that characterize the nondestructive photonic qubit detector (NPQD) are shown in Fig.\,3 in the main text together with numerical simulations. The theoretical model we employ is based on the cavity input-output theory \cite{Kuhn2015} used in a similar way as in \cite{Daiss2019}. The simulations were written in python together with the QuTiP package \cite{qutip}.

We consider weak coherent pulses in front of the qubit cavity \ket{\alpha} that are interacting with the atom-cavity system. After interaction, the photonic part might populate five different modes: the reflection at the resonator back into the fibre mode \ket{r_{0_\text{a},1_\text{a}}} and back into the fibre cladding due to an imperfect fibre-cavity mode matching \ket{r^o_{0_\text{a},1_\text{a}}}, the transmission through the resonator \ket{t_{0_\text{a},1_\text{a}}}, and the scattering and absorption losses given by the fibre-cavity mirrors \ket{m_{0_\text{a},1_\text{a}}} as well as by the atom \ket{a_{0_\text{a},1_\text{a}}}. The amplitude of the arising coherent fields in the just mentioned modes are given by:
\begin{align}
\begin{split}
r_{0_\text{a},1_\text{a}}&=\left(1-\mu_{FC}^2\frac{2\kappa_r}{N g^2/(i\Delta_a + \gamma)+i\Delta_c + \kappa}\right)\alpha, \\
r^o_{0_\text{a},1_\text{a}}&=\sqrt{1-\mu_{FC}^2}\mu_{FC}\frac{2\kappa_r}{N g^2/(i\Delta_a + \gamma)+i\Delta_c + \kappa}\alpha,\\
t_{0_\text{a},1_\text{a}}&=\mu_{FC}\frac{2\sqrt{\kappa_r\kappa_t}}{N g^2/(i\Delta_a + \gamma)+i\Delta_c + \kappa}\alpha,\\
m_{0_\text{a},1_\text{a}}&=\mu_{FC}\frac{2\sqrt{\kappa_r\kappa_m}}{N g^2/(i\Delta_a + \gamma)+i\Delta_c + \kappa}\alpha,\\
a_{0_\text{a},1_\text{a}}&=\mu_{FC}\frac{2g\sqrt{\kappa_r\gamma N}/(i\Delta_a+\gamma)}{N g^2/(i\Delta_a + \gamma)+i\Delta_c + \kappa}\alpha.
\end{split}
\end{align}
Here, $N$ is the number of atoms in the state that is coupled to the cavity mode (e.g. $N=0$ when the single atom is in state \ket{1_\text{a}} and $N=1$ when it is in \ket{0_\text{a}}). The atom has an atomic dipole decay rate $\gamma$ and a coupling strength with the qubit cavity mode of $g$. The spectral detuning between the incoming coherent field and the atomic transition (qubit cavity mode) frequency is given by $\Delta_a$ ($\Delta_c$). $\kappa_r$, $\kappa_t$ and $\kappa_m$ describe the field decay rates into the respective mode with a total field decay rate of $\kappa$. Additionally, we take into account the imperfect fibre-cavity mode matching given by $\mu_{FC}$. For convenience we simplify the notation of the loss modes to \ket{l_{0_\text{a},1_\text{a}}} $:=$ \ket{r^o_{0_\text{a},1_\text{a}}, t_{0_\text{a},1_\text{a}}, m_{0_\text{a},1_\text{a}}, a_{0_\text{a},1_\text{a}}}.
After the coherent pulse has interacted with the atom-cavity system, we can write the atom-photon state as
\begin{equation}
    \left|\Psi^1_{rla}\right> = \frac{\left|r_{0_\text{a}},l_{0_\text{a}}, 0_\text{a}\right>+
    \left|r_{1_\text{a}}, l_{1_\text{a}}, 1_\text{a}\right>}{\sqrt{2}}.
\end{equation}
A $\pi/2$ rotation operator is subsequently applied to the atomic part as the microwave pulse does in the experiment, leading to $\left|\Psi^2_{rla}\right>=\mathbb{1}_r\otimes\mathbb{1}_l\otimes \hat{R}_a(\pi/2)\left|\Psi^1_{rla}\right>$. With \ket{\Psi_{rla}} we calculate the parameters that characterize our NPQD which is described in the following paragraphs.

Unconditional nondestructive detection probability $P(0_\text{a})$: To investigate the detection probability $P(0_\text{a})$, we first trace out the photonic part in order to obtain the atomic part $\rho_a$. Then, we calculate the overlap with the photon detecting atomic state \ket{0_\text{a}}:
\begin{align}
\begin{split}
\rho_a &= \text{Tr}_{rl} \left|\Psi^2_{rla}\right>\left<\Psi^2_{rla}\right|,\\
p_0&=\left<0_\text{a}\right|\rho_a\left|0_\text{a}\right>,
\end{split}
\end{align}
which provides the probability to nondestructively detect the input qubit photon. Due to NPQD dark counts, there is the small probability $p_\text{DC}$ to detect the atom in state \ket{0_\text{a}} even if the NPQD did not interact with any photon. This is taken into account in order to obtain the final nondestructive detection probability
\begin{equation}
    P(0_\text{a}) = p_0 + (1-p_0)\cdot p_\text{DC}.
\label{eq:P0a}
\end{equation}
For an incoming coherent pulse with $\left|\alpha\right|^2\rightarrow0$, we expect $p_0$ to converge towards zero. However, eq.\,\ref{eq:P0a}
shows that $P(0_a)$ is then governed by $p_\text{DC}$ which sets the lower limit.

Nondestructive detection probability conditioned on qubit survival $P(0_\text{a}|\geq1_\text{oq})$:
This quantity is similarly evaluated as $P(0_\text{a})$ but additionally underlies the condition of a reflected photonic qubit. In the simulation we therefor project the reflection mode of state \ket{\Psi^1_{rla}} onto all non-vacuum components, using the projector operator $\hat{P}_r=\sum_{n_r>0}\left|n_r\right>\left<n_r\right|\otimes\mathbb{1}_l\otimes\mathbb{1}_a$. The atom-photon state after this projection becomes $\rho^1_{la}=\hat{P}_r \rho^1_{rla}\hat{P}_r/\text{Tr}(\hat{P}_r \rho^1_{rla}\hat{P}_r)$. In order to obtain the final atomic state, we consider the second atomic rotation $\rho^2_{la}=(\mathbb{1}_l\otimes \hat{R}_a(\pi/2))\rho^1_{la}(\mathbb{1}_l\otimes \hat{R}_a(\pi/2))^{\dagger}$ and trace out the photonic loss modes $\rho^2_{a}=\text{Tr}_l(\rho^2_{la})$. Additionally, we add the density matrix that was projected onto the photonic vacuum state and weight it with the probability of measuring a conventional detector dark count. We finally calculate the overlap between the resulting density matrix and state \ket{0_\text{a}}.

Mean photon number at the NPQD output conditioned on a nondestructive detection $\bar{n}_\text{oq}(0_\text{a})$:
To compute $\bar{n}_\text{oq}(0_\text{a})$, the atomic part of state \ket{\Psi^2_{rla}} is projected onto \ket{0_\text{a}}. To end up with the state in reflection mode $\rho_r(0_\text{a})$, the photonic loss modes \ket{l_{0_\text{a}}} are traced out. 
With this, we get $\bar{n}_\text{oq}(0_\text{a})=\left<\hat{n}_r\right>_{\rho_r(0_\text{a})}$ where $\hat{n}_r$ is the photon number operator acting on the state in reflection mode.
Furthermore, we include the contribution of false nondestructive detection events due to NPQD dark counts $p_\text{DC}$ by
\begin{equation}
    \tilde{n}_\text{oq}(0_\text{a})=\frac{\bar{n}_\text{oq}(0_\text{a})\cdot p_0+\bar{n}_\text{oq}(1_\text{a})\cdot(1-p_0)\cdot p_\text{DC}}{P(0_\text{a})},
\end{equation}
where $\tilde{n}_\text{oq}(0_\text{a})$ is governed by $p_\text{DC}$ as soon as $p_0$ converges towards zero whereas an absent $p_\text{DC}$ would lead to the convergence towards $\bar{n}_\text{oq}(0_\text{a})$.

Autocorrelation function $g^{(2)}(0)$ of the NPQD output field upon nondestructive detection:
Having $\rho_r(0_\text{a})$ derived, it is an easy task to calculate the well known second order correlation function,
\begin{equation}
    \tilde{g}^{(2)}(0) = \frac{\left<\hat{a}^{\dagger}_r\hat{a}^{\dagger}_r\hat{a}_r\hat{a}_r\right>_{\rho_r(0_\text{a})}}{\left<\hat{a}^{\dagger}_r\hat{a}_r\right>^2_{\rho_r(0_\text{a})}},
\end{equation}
where $\hat{a}_r$ is the photonic annihilation operator acting on the state in the reflection mode. In the limit of few detections for small $\left|\alpha\right|^2$, the dark counts of the SPDs result in an
uncorrelated signal. Therefor, we include the conventional detector dark counts by
\begin{equation}
    g^{(2)}(0)=\frac{g^{(2)}(0)\cdot p_{\geq2}+2\cdot p_{\geq1}\cdot p_\text{spd}+p^2_\text{spd}}{p_{\geq2}+2\cdot p_{\geq1}\cdot p_\text{spd}+p^2_\text{spd}},
\end{equation}
where $p_{\geq x}$ gives the integrated probability to find $\geq x$ photons in $\rho_r(0_\text{a})$. $p_\text{spd}$ describes the probability to measure a SPD dark count event.

\section{Distillation of photonic Fock state qubits}
An incoming single qubit photon would always imprint a $\pi$ phase shift on the phase of the atomic superposition state of our NPQD. However, we have characterized the NPQD with weak coherent pulses that are composed of odd and even photon number contributions. This situation causes an ambiguous phase shift that is described by $n\cdot\pi$ with $n$ as the photon number state. Since the NPQD signal heralds a $\pi$ phase shift, we conduct a projective parity measurement on the photon number state \cite{Daiss2019}. Moreover, in the range of $\left|\alpha\right|^2\ll1$ the nondestructive detection of an incoming qubit pulse leads to the distillation of an outgoing single photon that shows sub-Poissonian photon statistics, $g^{(2)}(0)<1$.

In the main text we provide a measurement of the second order correlation function of the reflected and nondestructively detected photonic pulses versus the mean photon number of incoming pulses. For this we have used the detection setup that consists of retardation waveplates, a polarizing beamsplitter and two SPDs (main text, Fig.\,1b), enabling the realization of a Hanbury Brown-Twiss setup \cite{HBT1956}.

As already mentioned, we observe $g^{(2)}(0)<1$ for $\left|\alpha\right|^2\lessapprox1$ (main text, Fig.\,3b). However, in the limit of $\left|\alpha\right|^2\rightarrow0$, the dark counts of the SPDs result in an uncorrelated signal that ultimately limits the lowest value obtained for $g^{(2)}(0)$. On the other side, $\left|\alpha\right|^2>1$, we have first expected a convergence for $g^{(2)}(0)$ towards one. However, unequal field reflection coefficients ($r_0$ and $r_1$), as we have discussed them in Methods, lead to a state after reflection $\propto$\ket{r_0\alpha}$ + $\ket{r_1\alpha} which allows for photon statistics with $g^{(2)}(0)>1$.

\printbibliography